\documentclass{article}
\usepackage[english]{babel}
\usepackage[utf8x]{inputenc}
\usepackage[T1]{fontenc}

\usepackage{fcds}
\usepackage{xcolor}
\usepackage[normalem]{ulem}

\usepackage[a4paper,top=3cm,bottom=2cm,left=3cm,right=3cm,marginparwidth=1.75cm]{geometry}

\usepackage{amsmath}
\usepackage{graphicx}
\usepackage[colorinlistoftodos]{todonotes}
\usepackage[colorlinks=true, allcolors=blue]{hyperref}

\newcommand{\mygraphic}[1]{\protect\includegraphics[height=#1]{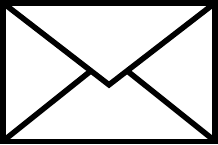}}
\newcommand{\myenv}{\mygraphic{.5em}}

\newcommand\blfootnote[1]{%
  \begingroup
  \renewcommand\thefootnote{}\footnote{#1}%
  \addtocounter{footnote}{-1}%
  \endgroup
}

\makeatletter
\renewcommand*{\@fnsymbol}[1]{\myenv}

\title{\textbf{Brilliant Challenges Optimization Problem Submission Contest\\Final Report}}
\author{Jan Badura$^{1,2,\myenv}$ \and Artur Laskowski$^{1,2, \myenv}$ \and Maciej Antczak$^{1,2,3}$ \and Jacek Blazewicz$^{1,2,3}$  \and Grzegorz Pawlak$^1$ \and Erwin Pesch$^{4,5}$ \and Thomas Villmann$^{6}$ \and Szymon Wasik$^{1}$}

\begin{document}
\maketitle

\blfootnote{
    \myenv \quad jan.badura@cs.put.poznan.pl
    
    \quad \myenv \quad artur.laskowski@cs.put.poznan.pl
}
\blfootnote{
    $^1$ \quad Institute of Computing Science, Poznan University of Technology
}
\blfootnote{
    $^2$ \quad European Center for Bioinformatics and Genomics, Poznan University of Technology
}
\blfootnote{
    $^3$ \quad Institute of Bioorganic Chemistry, Polish Academy of Sciences
}
\blfootnote{
    $^4$ \quad Center of Advanced Studies in Management, HHL Leipzig Graduate School of Management
}
\blfootnote{
    $^5$ \quad Department of Management Information Science, University of Siegen
}
\blfootnote{
    $^6$ \quad Computational Intelligence Group, University of Applied Sciences Mittweida\\[2ex]
}

\begin{abstract}
This paper concludes the Brilliant Challenges contest. Participants had to design interesting optimization problems and publish them using the Optil.io platform. It was the first widely-advertised contest in the area of operational research where the objective was to submit the problem definition instead of the algorithmic solutions. Thus, it is a crucial contribution to Open Science and the application of crowdsourcing methodology to solve discrete optimization problems. The paper briefly describes submitted problems, presents the winners, and discusses the contest's achievements and shortcomings. Finally, we define guidelines supporting the organization of contests of similar type in the future.
\end{abstract}


\section{Introduction}

\begin{figure}[ht]
\centering\includegraphics[width=\linewidth]{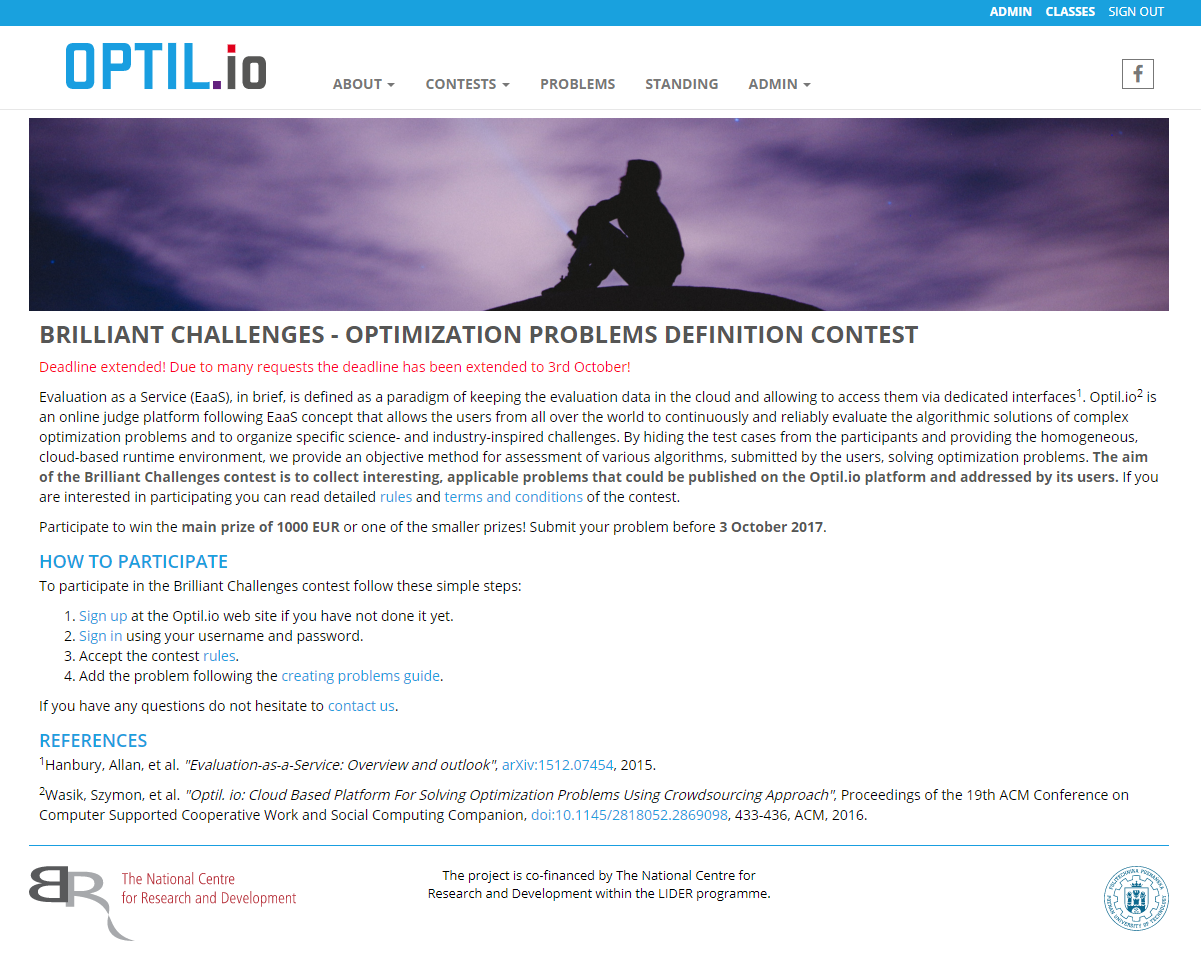}
\caption{Screen-shot presenting a home page of the Brilliant Challenges contest.}
\label{fig:homepage}
\end{figure}

Evaluation as a Service (EaaS), in brief, is defined as a paradigm of providing the evaluation data in the cloud via dedicated interfaces \cite{Hanbury_2015}. The  \href{www.optil.io}{Optil.io} \cite{Wasik_2016} is an online judge platform \cite{Wasik2018} following the EaaS concept \cite{Wasik_2017a}. It allows users from all over the world to continuously and reliably evaluate algorithmic solutions of complex optimization problems and organize specific science- and industry-inspired challenges. Hiding the evaluation test suite from the participants and providing a homogeneous, cloud-based runtime environment allows for reliable assessment of various algorithms, submitted by users, and efficiently solving optimization problems. The Brilliant Challenges contest aimed to collect interesting, applicable optimization problems that could have been published at the Optil.io platform and addressed by its users. This way, we also wanted to contribute to Open Science, especially in the context of Open Data \cite{Murray_2008}. We believe that scientific data should be made public whenever possible, to facilitate the research process. In the case of discrete optimization, those data consist not only of test instances but also algorithms implementation and evaluation environment. It is crucial to evaluate algorithms in terms of quality and efficiency developed by various researchers. The contest awards were funded by the research grant dedicated to young researchers. Therefore participants taking part in it had to be at most 37 years old. Detailed rules of the contest are presented in Appendix \ref{sec:rules}. A web page of the contest is presented in Figure \ref{fig:homepage}.

\section{Results}

During the contest, eight problems were submitted by researchers representing four different universities. The contest's initial interest was notably higher than that because it was advertised during the most significant international conference on Operational Research (OR) IFORS organized in Quebec, Canada. At the conference, ca. 50 researchers have declared that they are interested in participating in the contest or encouraging their Ph.D. students to do so. It was also advertised during the 12th Metaheuristics International Conference in Barcelona, Spain, and through various mailing lists related to the OR. Although the total number of problems submitted during the contest was below initial expectations, nevertheless, it allowed us to formulate many interesting conclusions (for details see Section \ref{sec:discussion}).

\subsection{Submitted problems}

\begin{table}[ht]
\centering
\begin{tabular}{p{2.5cm}|p{6cm}|r|r|r}
Name & Type & Inst. & Tot. time [s] & Freq. [min] \\\hline
\href{https://www.optil.io/optilion/problem/3013}{Ancient Ruins} & Orthogonal 2D unbounded knapsack problem & 61/0 & 2'690 & 180 \\
\href{https://www.optil.io/optilion/problem/3014}{Painting Artist} & 2D unbounded knapsack problem with irregular items, 2D pattern matching & 51/0 & 2'090 & 180 \\
\href{https://www.optil.io/optilion/problem/3015}{Short Chains} & Euclidean Steiner tree & 50/0 & 2'662 & 180\\
\href{https://www.optil.io/optilion/problem/3016}{Aerial Firefighters} & Cellular automaton simulation, maintenance scheduling  & 40/0 & 13'920 & 360\\
\href{https://www.optil.io/optilion/problem/3017}{3D Knapsack} & 3D orthogonal knapsack with rotations & 40/0 & 9'600 & 0\\
\href{https://www.optil.io/optilion/problem/3018}{Smart Customer} & Internet shopping optimization problem & 34/0 & 918 & 120\\
Sequential Partially Covering Problem & & N/A & N/A & N/A \\
FIFO Stack-Up Problem & & N/A & N/A & N/A 
\end{tabular}
\caption{\label{tab:problems}Problems submitted for the Brilliant Challenges contest. The \emph{Inst.} column presents the number of both public and private test instances submitted by the authors. The \emph{Tot. time} column represents the total processing time limit for all considered instances, i.e., the time required to evaluate the submitted solution to the problem if it utilizes the maximal processing time available for each instance. The \emph{Freq.} column presents a submission frequency, i.e., a minimal period of time that has to pass between two consecutive submissions. For two problems that were submitted in PDF the last three columns are not applicable. For the remaining problems their name is linked to the description at the Optil.io website.}
\end{table}

A summary of the submitted problems is presented in Table \ref{tab:problems}. The more detailed description is presented in the following sections. Descriptions presented in Sections \ref{sec:first-problem}-\ref{sec:last-problem} are only sketches of submitted problems. Their full description is published at the Optil.io website (links are provided in Table \ref{tab:problems}). According to the contest rules, participants were expected to submit their problems through the Optil.io platform using a dedicated form. Most of the authors followed those rules. However, two authors have submitted their problem descriptions by an e-mail, i.e.,  \emph{FIFO Stack-Up Problem} and \emph{Sequential Partially Covering Problem}. They are presented in the following subsections. However, as violating rules, they were not considered during the contest settlement. Moreover, the latter one was submitted by authors who had violated age restrained. They claimed to understand that they cannot be awarded, nevertheless being inspired by the contest, they wanted to submit their problem.

\subsubsection{Ancient Ruins}\label{sec:first-problem}\label{sec:ruins}

This is a variant of the \emph{orthogonal 2D unbounded knapsack problem} \cite{Han_2008}, i.e., problem of packing $n$ rectangular items of size $w_i \times h_i,1 \le i \le n$ with given values $v_i,1 \le i \le n$ into a single rectangular container of size $W \times H$ maximizing the total value of packed items. Here, we have an unlimited number of rectangular items (i.e., the problem is unbounded). Moreover, the problem considers rectangular items of any available size ($n=W \cdot H$). For each size of the rectangle $(w_i,h_j),1\le i,j\le n$ its value is defined as $v_{i,j}$. There is also one additional constraint that some given horizontal or vertical lines inside the container cannot belong to the interior of any considered item. The problem is supplemented with an interesting story about the treasure hunter who finds the map of an ancient castle and tries to compute the maximal budget required to build such a castle.

\subsubsection{Painting Artist}

This is a variant of \emph{2D unbounded knapsack problem with irregular items} \cite{DelValle_2012,Burke_2007} linked with \emph{2D pattern matching}. The problem's objective is to maximize the profit from placing $n$ irregular items on a rectangle of size $W \times H$. A subset of a rectangular grid defines the shape of any item. For each item, we know its value $v_i$, and there is an infinite number of its copies available (i.e., the problem is unbounded). Its worth to mention one additional constraint, i.e., some parts of the rectangular container are colored, and each pattern can only cover areas with the same color. Thus there is a need to implement a 2D pattern matching procedure for matching patterns to colored areas. The problem is supplemented with a story about a painter who wants to maximize some pictures' awesomeness.

\subsubsection{Short Chains}

It is an \emph{Euclidean Steiner tree} problem \cite{Korte_2001} with restricted areas. Restricted areas are defined as circles with a constant radius $R$. No Stainer point (i.e., an intermediate point that is not defined in the instance but can be added to decrease the total length of lines) can belong to any restricted area.

\subsubsection{Aerial Firefighters}\label{sec:fighters}

This problem is related to the simulation of a cellular automaton with von Neumann neighborhood \cite{vonNeumann_1951}. In this case, the automaton models how the fire spreads into neighboring cells leading to their damages \cite{Karafyllidis_1997}. Each cell $(i,j)$ is characterized by two attributes: a fire strength $s_{i,j}$ and the number of flammable resources $r_{i,j}$ that are initially equal to $r_{i,j}^0$ and decreased during the simulation when $s_{i,j}>0$. An objective is to minimize the amount of burned resources $R=\sum_{i,j} (r_{i,j}^0-r_{i,j})$ summed with a penalty for fields that were totally burned $P=\sum_{x,y} r_{x,y}^0$, where $r_{x,y}=0$. The score can be minimized by scheduling operations of planes that will drop water from the air. As such, this problem can be classified as a maintenance scheduling with tardiness minimization (the smaller tardiness, the fewer resources will burn) \cite{Cassady_2003}.

\subsubsection{3D Knapsack}\label{sec:knapsack}

This is a three-dimensional, orthogonal knapsack problem with rotations permitted around all three axes \cite{Diedrich_2008}. This problem was submitted as a purely scientific problem with neither a practical story nor additional constraints added.

\subsubsection{Smart Customer}\label{sec:shopping}

This is a variant of Internet shopping optimization problem \cite{Kovalyov_2010}. The problem assumes buying some set of items. Many sellers can sell each item, and each seller is characterized by various shipping costs that additionally depend on the weight of ordered items. The objective is to select a seller for each item to minimize the total cost of purchases and shipping costs. What differs this problem from the classic definition \cite{Kovalyov_2010} is an extension of an objective function, taking into account a component depending on the number of parcels that will be shipped (we also want to consider minimization of the number of parcels).

\subsubsection{Sequential Partially Covering Problem}

The sequential partially covering problem \cite{Nuriyeva_2016} is a subproblem of bandpass problem \cite{Babayev_2009}. It is an interesting problem occurring during the design of optical telecommunication networks. Its objective is to optimally pack information transmitted using different wavelengths into groups to obtain the highest cost reduction in optical communication networks' design and operation using wavelength-division multiplexing technology \cite{Kaminow_1996}. Unfortunately, the problem was submitted in PDF file only without test instances and the judge's source code for calculating the objective function's values. Thus it was not possible to integrate it with the Optil.io platform.

\subsubsection{FIFO Stack-Up Problem}\label{sec:last-problem}

This is a problem originated from logistics when there is a need to stack up bins from a set of conveyor belts onto pallets. Given is a list $Q = (q_1,... , q_k)$ of sequences of various types of products and a positive integer $p$. An objective is to decide whether there is a processing of $Q$, such that in each processing configuration of $ Q $, at most $p$ pallets are partially filled with products assuming that each pallet is dedicated to a single product type only \cite{Gurski_2015}. The problem was submitted by researchers inspired by the Brilliant Challenges contest. However, firstly, they were too old to participate in the contest, and secondly, they submitted only the problem description without either test instances or the judge's source code. It was an interesting addition to the contest. Nevertheless, it was not included in the final judgment.

\subsection{Problems analysis}

The diversity of submitted problems was significantly high. Authors submitted problems related to classic areas of combinatorial optimization (i.e., knapsack problem, Steiner tree) and its modern applications such as Internet shopping optimization and even problems that require computationally complex simulation to calculate the objective function's value (i.e., simulation of the cellular automaton). One problem was submitted in its classic form (i.e., 3D knapsack), but most of them have several additional constraints added. Both groups of problems are valuable for users of the Optil.io platform. The classical ones make it possible to compare new approaches with well-known and optimized algorithms in the field. In turn, elaborated problem versions usually have more practical applications.

The problems submitted by contestants use between 34 and 61 test instances to evaluate users' submissions. Usually, that number is high enough to compare submissions reliably. However, for some problems with a large number of constraints, such as Painting Artist, it could not be sufficient. No authors used the possibility to submit private instances that, in time-bounded challenges, ensure a reliable evaluation. This feature of the Optil.io platform allows authors to submit test instances dedicated only for the final evaluation purposes. Thus the value of the objective function computed for these instances is presented only to the author. It is beneficial to prevent over-fitting the solutions which are possible on publicly available test instances. 

A processing time limit for a single test instance evaluation varied from 30 to 360 seconds giving the total time required to evaluate the particular submission between 918 seconds (ca. 15 minutes) and 13920 seconds (ca. 4 hours). Even the latter limit is reasonable, as the Optil.io platform utilizes 40 engines evaluating submissions in parallel. To prevent a single user from submitting to many solutions one after another, the authors could define a time delay that has to elapse between consecutive submissions. Most of the problems have such a limit defined to the value between 2 and 6 hours. Such implementation should protect the Optil.io platform from being overloaded with too many submissions of the single user.

\subsection{Awards}

In this section, we present the prizes that have been awarded by the scientific committee of the contest based on the following criteria: scientific excellence, technical excellence, and an impact. All winning submissions were prepared to ensure a high technical level. They just needed minor corrections only. The opinion of the scientific committee about the remaining criteria for each winning submission is summarized below. Due to the protection of personal data, we present only author names for winning submissions. As can be easily noticed, most of the winning submissions were submitted by a single author (i.e., Sylwester Swat) who submitted high-quality solutions.

\subsubsection{First prize}

The first prize was awarded to the Ancient Ruins problem (see Section \ref{sec:ruins}) submitted by Sylwester Swat. Quality of the problem description is the highest, and the problem covers the most significant number of test instances, ensuring a reliable evaluation of submissions. Moreover, it is an exciting variant of the widely-known combinatorial optimization problem (i.e., 2D unbounded knapsack). Thus it has a high scientific impact.

\subsubsection{Second prize}

The second prize was awarded to the Aerial Firefighters problem (see Section \ref{sec:fighters}) submitted by Sylwester Swat. The problem description is a bit long and complicated. Nevertheless, it utilizes the critical aspect of simulating cellular automata. Recently, primarily due to the increase of available computing power, including high-performance GPU computing \cite{Millan_2017}, such a problem gains a significant interest. For this reason, its scientific impact was also classified as high. Moreover, its simulation-oriented character strengthened by maintenance scheduling stands out from the rest of the submitted problems.  

\subsubsection{Most entertaining}

The Ancient Ruins problem (see Section \ref{sec:ruins}) that won the first prize was also distinguished as the most entertaining problem. There were several reasons for this choice. This problem presents an exciting story which is not too long and is illustrated with helpful figures. Moreover, the input/output format passed to the submissions is quite simple. Thus the user of the Optil.io platform solving this problem will not have to spend much time parsing the data. Both scientific and technical committees were unanimous that many users will have much fun solving this problem.

\subsubsection{Industrial prize}

The industrial prize was awarded to the Smart Customer problem (see Section \ref{sec:shopping}) again submitted by Sylwester Swat. Awarding this prize to the only one submitted problem having a direct connection with the business activity was a natural choice. 

\subsubsection{Scientific prize}

A scientific prize was awarded to the 3D Knapsack problem (see Section \ref{sec:knapsack}) submitted by Artur Mostowski. It is a problem presenting the classical combinatorial optimization problem (i.e., 3D orthogonal knapsack with rotations). Thus it can be used to compare new approaches with many existing algorithms solving this type of problem \cite{Diedrich_2008,Zhao_2016}. For this reason, it was classified as a problem with the highest scientific impact. The only shortcoming of the problem is the set of test instances used, which did not include examples from standardized instances library that has already been published.

\section{Discussion}\label{sec:discussion}
Here, we summarize the main achievements of the Brilliant Challenges contest and the improvements implemented in the Optil.io platform due to the feedback we have received. Next, we point out the drawbacks and limitations identified during the contest. Finally, the potential reasons for the rather low number of submissions were also discussed.

\subsection{Achievements}
\label{sec:achievements}
A variety of original problems were submitted during the contest. Most of them were carefully prepared with exciting and clear descriptions supported by clarifying figures. One author tried to make the problem more attractive to users by including the logo of the Optil.io platform and the problem name inside the test data. Most of the problems had enough test instances to provide a reliable evaluation of solutions, which shows that most authors understand the necessity to correctly evaluate users' solutions. However, we observed a lack of private instances. We addressed this problem in section \ref{sec:shortcomings}. 

As a result of constructive feedback from contestants during the contest, we identified some features that had to be implemented. The most important one concerned the extension of the upload form for test instances. Previously there was one specific way to upload a set of instances as a ZIP file. Now we changed it so that the instance name may be any string ending with 'in' extension (if the output is known, the corresponding file with 'out' extension may be introduced). This change simplifies the instances management in the Optil.io.

The second notable improvement allows authors to submit solutions to their problems before making the problem public. Initially, the judge program proposed for the problem had to be verified by Optil.io administrator since it was run with relatively high privileges (i.e., no time limit). This change enabled authors to test Optil.io workflow efficiently.

Last but not least, the achievement of Brilliant Challenges is the new popularity and user base that the Optil.io platform gained. Some of the participants showed interest in organizing optimization contest as it is a convenient way to publish public instances during the contest and private instances after it ends. It is worth to underline that since 2018, \textit{The Parameterized Algorithms and Computational Experiments Challenge} was evaluated using the Optil.io platform \cite{bonnet2018pace,dzulfikar2019pace}.


\subsection{Shortcomings}
\label{sec:shortcomings}
The contest showed that for many researchers, it is difficult to prepare test instances and evaluation programs. None of the participants prepared private instances for their problem. We believe that neither the rules nor instance upload interface emphasized the difference between public and private instances and how they are used in Optil.io workflow. Therefore the authors omitted them, as the difference is visible only from a problem-solver perspective. Creating a separate set of private instances is important so that the problem solvers do not overfit their solutions for some known set of instances.

Most of the authors had problems with implementing a proper judge that computes the objective function value. Usually, computing this value is not a problem, but handling incorrect solutions is a big challenge. To handle them correctly, the judge has to detect all possible errors in the generated output and provide an informative message to the user of the Optil.io platform. Here, errors are related to the incorrect format of the output data (i.e., formatting inconsistent with the format presented in the problem description) and solutions that do not satisfy all criteria required in the problem definition. Many judges simply terminate with runtime error instead of ensuring valuable errors handling and notifying the Optil.io platform. The evaluation result often used the incorrect format, even though a template of a supported judge program was provided. Each submission had to be tuned by somebody experienced in using the Optil.io system and evaluation procedure of algorithmic solutions to optimization problems.

\subsection{Reasons for low interest}
\label{sec:resonsforlowinterest}
We tried to identify the cause of the low number of submissions as opposed to high initial interest. One of the main reasons seems to be the technical difficulties related to the correct implementation of the judge program. The other reason is that the researchers do not see the benefits of publishing evaluation data (excluding evaluation data for quite basic variants of very well-known problems). They also do not want to spend time on publishing evaluation data following the open science guidelines. They prefer to solve another problem at that time. 

\section{Conclusions}
Organizers and participants of the Brilliant Challenges contest acknowledged it as successful. Most of the submitted problems were carefully prepared to fulfill precisely the requirements imposed in the contest rules. Both the scientific and technical excellence of the winners is also worth to underline. 
The Optil.io platform greatly benefits from the contest. Following the contest announcement, we observed tens of newly created accounts and hundreds of algorithmic solutions submitted.
We hope the Optil.io platform would be often applied by the scientific community to reliably evaluate solutions of science- and industry-inspired optimization problems submitted by practitioners.

It was the first widely-advertised challenge in the area of operational research, in which the objective was \textbf{not} to submit algorithmic solutions for some problem, but to publish the problem itself. The challenge proved that the application of crowdsourcing could be very successful in defining interesting problems \cite{Wasik_2018b} in the OR area. For this reason, we propose guidelines supporting organization of such type of contests in the future:
\begin{enumerate}
    \item \textbf{Precisely describe what has to be submitted and validate it automatically.} We tried to make all descriptions as clear and precise as possible, but still, all researchers missed the need to define private instances. It could be validated automatically but was not. The author of a similar challenge should keep in mind that the process of publishing the problem is much more difficult than just implementing the algorithm solving it to ensure the participant will be carefully guided during the process.
    \item \textbf{Explain how submitted problems will be used.} It was also missed. Probably, if we explained better, in the Challenge rules, how the defined problems are process in the Optil.io platform and what evaluation methodology is applied, it would help participants to include all required elements, such as private instances, in their submissions.
    \item \textbf{Provide active continues support to the participants.} As mentioned before, defining the problem itself is complex, and most of the researchers in the area of operational research do not have experience in administrating online judge platforms. For this reason, the support to the participants should be active, i.e., provided not only when they ask for help, but also initiated by organizers who should track submissions uploaded by them. It should also be continuous, i.e., provided actively during the whole challenge and not only after its end.
    \item \textbf{Provide user-friendly forms to upload all required description elements.} The fact that the Optil.io system allowed users to upload all required data easily, i.e., test instances, evaluation script, and problem description, helped to get high-quality submissions.
    \item \textbf{Advertise the challenge well and fund some prizes.} Researchers are not always aware of the benefit of publicly publishing the detailed problem definition and its evaluation test instances. It is worth advertising a challenge during a few widely-known conferences and funds some prizes to encourage them to do so. Those could be financial rewards and the possibility to be the co-author of some publication summarizing the results of the challenge published in the high-impact journal.
    \item \textbf{Provide templates wherever possible.} We provided a code of sample evaluation script and a template of the problem description during the contest. Many submissions benefitted from it; participants to prepare high-quality problem definitions because they could focus on the content instead of its presentation.
\end{enumerate}
We believe that the aforementioned guidelines could also be valuable for other researchers dealing with Open Science in operational research and discrete optimization. Some of them may seem obvious, but they can be used as a to-do checklist crucial during the organization of similar challenges.

\bibliographystyle{acm}
\bibliography{brilliant}

\appendix
\section{Contest rules}\label{sec:rules}

\begin{enumerate}
\item
  Each participant is responsible for submitting a novel optimization problem
  (having intellectual property rights to it) through the Optil.io
  platform together with test instances and judge program following the
  \href{https://www.optil.io/optilion/help/create}{creating problems
  guide}. The submission deadline is \textbf{3rd October 2017}
  regardless of the user time zone.
\item
  The results of the contest will be announced on the Optil.io website
  on \textbf{15th October 2017}.
\item
  Every participant that submits the problem during the contest will be
  included in a final ranking. However, only contestants who are
  \textbf{between 18 and 37 years old} on 1st January 2017 are eligible
  for winning monetary prizes. The prize will be paid as payment for
  the intellectual property rights to the problem description.
\item
  Each problem has to be submitted by a single participant (called
  the corresponding author), who will receive a monetary prize in case of
  winning the contest. A team and all of the authors can design the problem can be included in the problem's description by setting the
  authors list field. Moreover, all of them will be included in the
  final certificate. However, only the corresponding author of the team
  that submitted this problem is responsible for the distribution of the
  potential prize among the team.
\item
  The scientific committee will award five prizes:

  \begin{itemize}
  \item
    1st prize (\textbf{1000 EUR}).
  \item
    2nd prize (\textbf{700 EUR}).
  \item
    Special distinction for the most entertaining problem (\textbf{700
    EUR}).
  \item
    Distinction for the most interesting industry-inspired problem
    (\textbf{400 EUR}).
  \item
    Distinction for the most interesting science-inspired problem
    (\textbf{400 EUR}).
  \end{itemize}
\item
  The scientific committee will assess all submitted problems
  based on the following criteria:

  \begin{itemize}
  \item
    Scientific excellence - the quality of the submitted problem
    description and test instances.
  \item
    Technical excellence - the quality of the submitted judge program
    and compliance with the guidelines used for preparing problems
    available in the
    \href{https://www.optil.io/optilion/help/create}{creating problems
    guide}.
  \item
    Impact - scientific and industrial importance of the problem.
  \end{itemize}
\item
   The special distinction for the most entertaining problem will be awarded to the problem that will be assessed by the technical committee as the most interesting problem for users of the Optil.io platform. Factors that can make the problem interesting are well-written problem descriptions, the short and clear format of input and output, and ease in preparing the first solution.
\item
  Distinctions for the most interesting problems will be awarded by the scientific committee based on the importance of the problem for
  industry and science, respectively.
\item
  To participate in the contest one has to:

  \begin{itemize}
  \item
    \href{https://www.optil.io/optilion/user/create}{Create an account}
    on the Optil.io platform.
  \item
    Accept the rules that govern this contest using the online form. As
    a result of that acceptation, the user will receive additional
    privileges for authoring new problems.
  \item
    Create at least one problem. Each participant can submit many
    problems and each of them will be assessed in the contest.
  \item
    The participant should attach a short, one-page report describing
    why the problem is important and how test instances were selected.
    The report should be written in plain text and pasted into the dedicated field that is displayed when the problem is sent for
    moderation.
  \end{itemize}
\item
  By accepting these rules, the participant accepts also
  \href{https://www.optil.io/optilion/brilliant/terms}{Terms and
  Conditions} of the contest.
\item
  Any questions should be sent using the
  \href{https://www.optil.io/optilion/contact}{contact us} page.
\item
  The scientific and technical committees consist of:

  \begin{itemize}
  \item
    Dr hab. Szymon Wasik, Poznan University of Technology (Chief organizer)
  \item
    Prof. Jacek Blazewicz, Vce Director of IFORS (Head of Scientific
    Committee)
  \item
    Prof. Erwin Pesch, Siegen University (Member of Scientific
    Committee)
  \item
    Prof. Thomas Villmann, University of Applied Sciences Mittweida
    (Member of Scientific Committee)
  \item
    Prof. Grzegorz Pawlak, Computing Science Professor, and Entrepreneur
    (Member of Scientific Committee)
  \item
    Dr hab. Maciej Antczak, Poznan University of Technology (Head of
    Technical Committee)
  \item
    Jan Badura, Poznan University of Technology (Member of Technical
    Committee)
  \item
    Artur Laskowski, Poznan University of Technology (Member of
    Technical Committee)
  \end{itemize}
\end{enumerate}

\end{document}